\documentclass[aps,prd,nofootinbib,twocolumn,superscriptaddress,floatfix,notitlepage]{revtex4-1}

\usepackage{graphicx}
\usepackage{amsmath,amsfonts,amssymb}
\usepackage{color}
\usepackage[breaklinks,colorlinks,urlcolor=blue,citecolor=blue,linkcolor=magenta]{hyperref}
\usepackage{verbatim}

\newcommand{\ie}{i.e.\,}

\newcommand{\be}{\begin{equation}} 
\newcommand{\ee}{\end{equation}}
\newcommand{\bea}{\begin{equation}\begin{aligned}} 
\newcommand{\eea}{\end{aligned}\end{equation}}

\newcommand{\td}{{\rm d}}

\begin{document}

\title{Bubble dynamics in fluids with $N$-body simulations}

\author{Marek Lewicki}
\email{marek.lewicki@fuw.edu.pl}
\affiliation{Faculty of Physics, University of Warsaw ul.\ Pasteura 5, 02-093 Warsaw, Poland}
\author{Ville Vaskonen}
\email{vvaskonen@ifae.es}
\affiliation{Institut de Fisica d'Altes Energies, Campus UAB, 08193 Bellaterra (Barcelona), Spain}
\author{Hardi Veerm\"ae}
\email{hardi.veermae@cern.ch}
\affiliation{Keemilise ja bioloogilise f\"u\"usika instituut, R\"avala pst. 10, 10143 Tallinn, Estonia}

\begin{abstract}
We present a new approach to studies of bubble dynamics in fluids. Relying on particle-based simulations, this method is general and suitable for cases where the commonly used perfect fluid description fails. We study expanding true vacuum bubbles surrounded by free or self-interacting particles and quantify how self-interactions affect the terminal bubble wall velocity. We find that, for sufficiently strongly self-interacting fluids, local thermal equilibrium is maintained around the bubble wall and the fluid profile is similar to that obtained with the perfect fluid description.
\end{abstract}

\maketitle

\section{Introduction}

The detection of gravitational waves (GWs) by LIGO~\cite{LIGOScientific:2016aoc} started a new era in astrophysics and cosmology. With many new GW experiments planned in the coming decades~\cite{Punturo:2010zz,Hild:2010id,Janssen:2014dka,Graham:2016plp,Audley:2017drz,Graham:2017pmn,LISA:2017pwj,Badurina:2019hst,Bertoldi:2019tck,Badurina:2021rgt} the potential of probing the early Universe through searches of stochastic GW backgrounds will increase tremendously. In fact, pulsar timing experiments have recently been reporting possible hints of a stochastic background at very low frequencies~\cite{NANOGrav:2020bcs,Goncharov:2021oub,Chen:2021rqp,Antoniadis:2022pcn} that might have originated from the early Universe~\cite{Ellis:2020ena,Blasi:2020mfx,Vaskonen:2020lbd,DeLuca:2020agl,Nakai:2020oit,Ratzinger:2020koh,Kohri:2020qqd,Vagnozzi:2020gtf,Neronov:2020qrl,Middleton:2020asl,Samanta:2020cdk}. We focus on first order phase transitions that are a common phenomenon in particle physics models and a possible GW source~\cite{Caprini:2015zlo,Caprini:2019egz,LISACosmologyWorkingGroup:2022jok}.

First order phase transitions are characterized by the nucleation of bubbles of the broken phase in the symmetric phase background~\cite{Coleman:1977py,Callan:1977pt,Linde:1981zj}. These bubbles then expand until they collide and convert the entire Universe to the broken phase. Interaction of the bubble walls with the ambient fluid is among the key topics currently under investigation by the community as it is key in determining the terminal velocity of the wall. Crucially, whether the walls reach a steady state before colliding dictates if the GW signal is primarily sourced by bubble collisions~\cite{Kosowsky:1992vn,Cutting:2018tjt,Ellis:2019oqb,Lewicki:2019gmv,Cutting:2020nla,Lewicki:2020jiv,Giese:2020znk,Ellis:2020nnr,Lewicki:2020azd} or fluid related sources~\cite{Kamionkowski:1993fg,Hindmarsh:2015qta,Hindmarsh:2016lnk,Hindmarsh:2017gnf,Ellis:2018mja,Cutting:2019zws,Hindmarsh:2019phv,Ellis:2020awk}. In the latter case, accurate modeling of the fluid motions induced by the growing bubbles is important for precise GW spectrum estimates~\cite{Hindmarsh:2019phv,Ellis:2020awk,Jinno:2020eqg,Gowling:2021gcy}. The wall velocity is crucial also for the possible generation of baryon asymmetry during the transition~\cite{Kuzmin:1985mm,Cohen:1993nk,Rubakov:1996vz, Morrissey:2012db,Cline:2020jre,Cline:2021iff,Lewicki:2021pgr}.

Several approaches have been developed to compute the wall velocity in particle physics models. For relatively weak transitions, the starting point is local thermal equilibrium~\cite{Espinosa:2010hh,Konstandin:2010dm,BarrosoMancha:2020fay,Balaji:2020yrx,Giese:2020rtr,Ai:2021kak,Wang:2022txy}, and accurate estimates also include perturbations around the equilibrium~\cite{Liu:1992tn,Turok:1992jp,Dine:1992wr} found by solving a system of Boltzmann transport equations~\cite{Moore:1995ua,Moore:1995si,Konstandin:2014zta,Kozaczuk:2015owa,Dorsch:2018pat,Laurent:2020gpg,Friedlander:2020tnq,Cline:2021iff,Lewicki:2021pgr,Laurent:2022jrs}. However, the result can be approximated assuming equilibrium, if the wall reaches a steady state with a shell of heated fluid around the bubble~\cite{Lewicki:2021pgr,Laurent:2022jrs}. If the wall accelerates beyond that point and only the fluid inside the bubble is heated, the solution describes a detonation~\cite{Steinhardt:1981ct}. In this case, the friction drops to the point where we reach large wall velocities for which perturbations become large, and a different computational method is necessary. This brings us to strong transitions where the wall is thin compared to the wavelength of particles in the fluid, and interactions can be treated in terms of transmission and reflection coefficients in the WKB approach~\cite{Arnold:1993wc}. The leading order result predicts a friction term proportional to the mass difference between particles on each side of the wall~\cite{Bodeker:2009qy}. For ultrarelativistic walls, emission of soft gauge bosons upon wall crossing becomes important~\cite{Bodeker:2017cim}. The scaling of friction with velocity is still a matter of some debate~\cite{Azatov:2020ufh, Hoche:2020ysm,Gouttenoire:2021kjv}\footnote{Holographic methods~\cite{Maldacena:1997re} have been employed to compute the wall velocity in strongly coupled theories~\cite{Bea:2021zsu,Bigazzi:2021ucw,Bea:2022mfb}}.

Dynamical simulations of the fluid in local thermal equilibrium have been conducted with an effective coupling between the scalar field and the fluid~\cite{Ignatius:1993qn,Kurki-Suonio:1995yaf}. These results were crucial for determining the GW spectrum~\cite{Hindmarsh:2015qta,Hindmarsh:2017gnf,Cutting:2019zws} produced by the fluid motion after the transition and the fate of its remnants, such as heated fluid droplets still in the old symmetric phase~\cite{Kurki-Suonio:1995yaf,Cutting:2022zgd}.
However, with this modeling, the connection to particle physics is unclear. The effective coupling needs to be traded for the wall velocity, which has to be computed beforehand and depends on the fluid profile around the wall.

In this work, we propose a novel dynamical approach for studying the coupled wall-plasma system. As an application, by starting with physical parameters, we compute the wall velocity and find the fluid profiles. Our description is easily applied to non-equilibrium cases~\cite{Baker:2019ndr,Baldes:2020kam,Baker:2021sno,Baker:2021zsf}. We model the wall as a phase boundary with energy conservation determining whether particles will penetrate the wall or be reflected.

\section{Theoretical background}

\subsection{Particle--wall interactions}

We consider a classical system of point-particles whose mass depends on the value of a scalar field $\phi$. For generality, we start with a generic curved background. Omitting particle self-interactions, the action of such systems is
\bea \label{eq:S_FVB+PP}
    S =& 
    \int \td^4 x \, \sqrt{-g} \left[\frac{1}{2}(\partial \phi)^2 - V(\phi)\right] \\
    &- \sum_n \int \td \tau \, m(\phi(x^{\mu}_n)) \sqrt{g_{\mu \nu}{x'}_n^\mu {x'}_n^\nu}\, ,
\eea 
where ${x'}^\mu \equiv \td x^\mu/\td \tau$, $\tau$ parametrizes the trajectories, and $n$ labels the particles. The scalar field obeys\footnote{The particle density is $n(x) \equiv |g|^{-1/2} \sum_n \delta^3(x^{i} - x^{i}_n(t)) = \int \frac{\td^3 p}{(2\pi)^3} f(x,p)$.}
\bea\label{eq:eom_FVB+PP}
    \Box \phi + V' 
&    = - \sum_n \, \frac{\dot s_n}{\sqrt{g}}\delta^3(x^{i} - x^{i}_n(t)) \partial_\phi m(\phi) 
\\ 
&    = - \int \frac{\td^3 p}{(2\pi)^3 2E} f(x,p) \partial_\phi m^2(\phi) \, ,
\eea
where dot denotes derivative with respect to $t$, $\dot s_n \equiv \sqrt{g_{\mu \nu}{\dot x}_n^\mu {\dot x}_n^\nu}$ and $f(x,p)$ is the particles phase space distribution. In equilibrium, the right hand side of Eq.~\eqref{eq:eom_FVB+PP} is the derivative of the thermal contribution to the scalar field effective potential~\cite{Dolan:1973qd,Moore:1995si}. In our approach, the fluid is described as a collection of particles, thus the corrections to $V$ are explicitly accounted for by particle-wall interactions.

The action~\eqref{eq:S_FVB+PP} does not contain interactions between particles. The Boltzmann equation, describing an interacting gas, has the general form $\mathbf{L}(f) = \mathbf{C}(f)$, where $\mathbf{L}$ and $\mathbf{C}$, denote the Liouville and the collision operator, respectively~\cite{Kolb:1990vq}. The Liouville operator is linear in $f$, describes the motion of non-interacting particles, including interactions with the wall, and is fully determined by the action~\eqref{eq:S_FVB+PP}. The non-linear collision operator is responsible for maintaining thermal equilibrium and the responsible particle self-interactions must be implemented separately. This will be addressed in Sec.~\ref{sec:sim}. We will consider both free and self-interacting particles.

By the action~\eqref{eq:S_FVB+PP}, the equations of motion of particles are $\left(m {x'}^{\mu} \right)' + m \Gamma^{\mu}_{\rho\sigma} {x'}^\rho {x'}^\sigma - \partial^\mu m = 0$, where $\Gamma^{\mu}_{\rho\sigma}$ denotes the Christoffel symbols, and reduce to the geodesic equations with a constant $m$. In terms of the canonical momenta $p_{\mu} \equiv -\partial L/\partial {x'}^{\mu}$, they can be expressed as ${p'}^{\mu} + \Gamma^{\mu}{}_{\rho\sigma} p^\rho {x'}^\sigma = \partial^\mu m$ and imply the on-shell condition $\left(p^2 - m^2 \right)' = 0$. 

Consider now a scalar field bubble. We assume that the interactions of the particles with the bubble wall are sufficiently localized in spacetime so that curvature can be neglected, and that the wall profile does not change during particle-wall interactions. This amounts to working in Minkowski spacetime, where we recover the usual 4-momentum $p_{\mu}  = m (\gamma, \gamma \dot x_{i})$ with $\gamma = 1/\sqrt{1-v^2}$. In the bubble wall frame and the $x$-axis normal to the wall the equations of motion simplify to $\dot E = \dot{p}_y = \dot{p}_z = 0$ and $\dot p_x = -\gamma^{-1}\partial_x m$. The change in 4-momentum is thus completely characterised by $p_x^2 + m^2 = {\rm constant}$. 

Consider the wall centered at $x=0$, assume that $m(x)$ approaches asymptotically a constant, $m_{\pm} = m(\pm \infty)$ and choose $m_{+} > m_{-}$. A particle coming from $x<0$ towards $x>0$ can either penetrate the wall if $E > m_+$ or get reflected from the wall, $p_x \to -p_x$, if $E < m_+$. In the former case, the momentum of the particle decreases as $p_x^2 \to p_x^2 + m_{-}^2 - m_{+}^2$. Since $m_{+} > m_{-}$, a particle coming from $x>0$ towards $x<0$ always penetrates the wall and its momentum increases as $p_x^2 \to p_x^2 + m_{+}^2 - m_{-}^2$.

Interactions with a moving boundary can be derived by boosting the above results to the fluid frame. We can use the walls normal $n^{\mu}$ to express the change of the particle momentum in a manifestly covariant way,
\be\label{eq:delta_p}
    p^\mu \to p^\mu + n^{\mu} \, n{\cdot} p \, {\cal F}(-n{\cdot} p) \,,
\ee
where the function ${\cal F}$ depends on the direction from which the particle approaches the wall. For particles approaching the wall from the $m_-$ region, it is given by
\be\label{eq:F-}
    {\cal F}_{-}(u) \equiv
\begin{cases}
    2 \,, & 0 < u < \Delta m \,, \\
    1 - \sqrt{1 - \frac{\Delta m^2 }{u^2}} \,, & u \geq \Delta m \,,
\end{cases}
\ee
where $\Delta m^2 \equiv m_{+}^2 - m_{-}^2$, and for particles approaching the wall from the $m_+$ region by
\be\label{eq:F+}
    {\cal F}_{+}(u) \equiv \theta(-u)\left(1 - \sqrt{1 + \frac{\Delta m^2 }{u^2}}\right) \,,
\ee
where $\theta$ denotes the step function. We fix $n^2 = -1$ and the sign of $n^\mu$ by $n_{\mu} \propto \partial_\mu m$ so that the normal is directed towards mass growth. For example, for a spherical wall enclosing the $m_{-}$ region moving at velocity $v_w$, the non-zero components of $n^\mu$ are $n^t = v_w \gamma_w$ and $n^r = \gamma_w$.

The number of collisions from the $+$ and $-$ regions per time interval $\td t$ on a surface element $\td S$ is
\be
    \frac{\td^2 N_{\pm}}{\td t \,\td S} 
    = \mp \int \frac{\td^3 p}{(2\pi)^3} f_{\pm}(p) \theta(\mp v_{\rm rel}) v_{\rm rel} \,,
\ee
where $v_{\rm rel} = -n{\cdot} p/(E\gamma_w)$ is the relative velocity between the particle and the wall in the direction of $\vec n$, and $f_{\pm}(p)$ denote momentum distributions near the wall. By Eq.~\eqref{eq:delta_p}, the energy transferred in each collision is $n_0\, n{\cdot} p {\cal F}(-n{\cdot} p)$. The pressure difference across the bubble wall caused by particle collisions is determined by energy transfer $\delta E$ from the bubble to the particles as the bubble's volume changes by $\delta V$,
\bea\label{eq:P}
    \Delta P 
&    \equiv \frac{\delta E}{\delta V} 
   = \int \frac{\td^3 p}{(2\pi)^3}\sum_{i \in \pm}  f_{i}(p)\frac{(n{\cdot} p)^2}{E_i}{\cal F}_{i}(-n{\cdot} p)  \, ,
\eea
where $E_{\pm}^2 = m_{\pm}^2 + p^2$. We will assume a spherical bubble and isotropic $f_{\pm}(p)$ so that the pressure is uniform across the wall and $\Delta P$ is independent of $n^{\mu}$.

In the limit $|v_w| \to 1$ or, equivalently, $(n{\cdot} p)^2 \to \infty$, one of the ${\cal F}_{\pm}$ coefficients always vanishes because $n{\cdot} p > 0$ when $v_w \to 1$ and $n{\cdot} p < 0$ when $v_w \to -1$, and we find
\be
    \lim_{v_w \to \pm 1} \Delta P
    = \pm \Delta m^2 
    \int \frac{\td^3 p}{(2\pi)^3} \frac{f_{\mp}(p)}{2E_\mp} \,,
\ee
agreeing with the pressure difference found in Ref.~\cite{Bodeker:2009qy}. For example, assuming a relativistic species with particle density $n$ that follows Maxwell-Boltzmann distribution, we find that $\Delta P = \pm \Delta m^2 n_{\mp}/(4T_{\mp})$. However, our result~\eqref{eq:P} is more general and allows us to estimate the terminal bubble wall velocity.

\subsection{Bubble dynamics}
\label{sec:bubble}

Consider an $O(3)$ symmetric scalar field configuration in which the field interpolates between two minima of its potential separated by a potential energy difference $\Delta V$ chosen such that outside the bubble $V=0$ and inside $V= - \Delta V < 0$. In the thin--wall limit, the scalar field action of the bubble in vacuum is~\cite{Darme:2017wvu,Ellis:2019oqb}
\be
    S_\phi  
    = \int \td t \left[ -4\pi \sigma R^2 \sqrt{1-\dot R^2} + \frac{4\pi}{3} R^3 \Delta V \right] \,,
\ee
where $R$ denotes the bubble radius and $\sigma \equiv \int \td \phi \sqrt{2 V}$ is the surface tension. Correspondingly, the energy of the bubble is
\be
    E_\phi 
    = 4\pi \sigma \frac{R^2}{\sqrt{1-\dot R^2}} - \frac{4\pi}{3} R^3 \Delta V \,.
\ee
In vacuum, the bubble's energy is conserved and the equation of motion for the bubble radius follows from $\dot E_\phi=0$.

To include the effect of interactions with the surrounding particles, we assume that the total energy of the system is conserved. Thus, $R$ obeys
\be \label{eq:Reom}
    \ddot R + 2\frac{1-\dot R^2}{R} 
    = \frac{(1-\dot R^2)^{3/2}}{\sigma} \left(\Delta V - \Delta P \right)\,,
\ee
where we used energy conservation to write $\dot E_\phi/4\pi \dot{R} R^2 = -\Delta P$, with $\Delta P$ given in Eq.~\eqref{eq:P}.

We consider expanding true vacuum bubbles relevant for cosmological phase transitions\footnote{We consider false vacuum bubbles relevant for collapse of the last false vacuum regions in phase transitions~\cite{Kodama:1982sf,Kurki-Suonio:1995yaf,Baker:2021nyl,Cutting:2022zgd} or false vacuum patches of inflationary origin~\cite{Garriga:2015fdk,Deng:2017uwc,Maeso:2021xvl} in a separate study~\cite{Lewicki:prep}.} and a single particle species. The bubble expansion is driven by the potential energy difference $\Delta V > 0$ and, because the particles are heavier inside the bubble than outside, the pressure $\Delta P$ is positive and increases with the bubble wall velocity $v_w$. The bubble wall asymptotically reaches a terminal velocity, determined by how fast $\Delta P$ increases with $v_w$, if $\Delta P$ asymptotically reaches $\Delta V$.

We consider scenarios in which particles exist only outside the bubble with the Maxwell-Boltzmann distribution, $f_-(p) = e^{-E_-/T_-}$, and $m_- \ll T_-$. The system is characterised by the temperature far outside of the bubble $T_-/m_+$, the mass ratio $m_-/m_+$, and the strength of the transition, 
\be\label{eq:alpha}
    \alpha \equiv \frac{\Delta V_T}{\rho_-} \,, \quad 
    \Delta V_T \equiv \Delta V - T \int \!\frac{{\rm d}^3 p}{(2\pi)^3} f_-(p) \,,
\ee
where $\rho_-$ denotes the energy density of the fluid far outside of the bubble and $\Delta V_T$ the potential energy difference including the thermal corrections integrated from the last term in Eq.~\eqref{eq:eom_FVB+PP} for the Maxwell-Boltzmann distribution. By Eq.~\eqref{eq:Reom} with $\Delta V - \Delta P = \Delta V_T$, the surface tension $\sigma$ determines the critical radius $R_c = 2\sigma/\Delta V_T$. If $R > R_c$, the bubble expands.

\section{Simulation set-up}
\label{sec:sim}

We study the dynamics of the coupled system consisting of the fluid and a scalar field bubble with $N$-body simulations. The simulation box, as which we consider a cube with periodic boundary conditions, is initialized in such a way that it includes the thin-wall bubble and particles inside and outside the bubble with the chosen number densities. The positions of the particles are chosen randomly from an uniform distribution, and their velocities are picked from the chosen momentum distribution. 

To include particle self-interactions, we consider a simple model of elastic $2\to 2$ scatterings based on hard spheres. The self-interaction strength is characterized by the interaction radius $r_c$, and the collision is assumed to occur when the two particles are within distance $2r_c$ in the center of mass (CM) frame, that is, when 
\be
    (\hat p \times \delta \vec x)^2 + \gamma^{2} (\hat p\cdot \delta \vec x)^2 \leq (2r_c)^2\, ,
\ee
where $\pm\hat p$ are the directions of 3-momenta of the particles, $\delta \vec x$ denotes their separation, and $\gamma$ is the CM-frame Lorentz factor that accounts for length contraction. To obtain the Maxwell-Boltzmann distribution as the system reaches equilibrium, the collisions must occur with a probability $\propto \sqrt{E_1 E_2}$, when the collision criterion is satisfied within a timestep. We take the CM scattering angle to be random after the collision.

The simulation proceeds in time step-by-step and at each time step the following operations are carried out: 
\begin{enumerate}
\item The positions of particles are evolved using the Euler method and their collisions with the wall are checked. If the particle collides with the wall, the timestep is split so that, first, the particle is evolved to the bubble wall, then its momentum is changed according to Eq.~\eqref{eq:delta_p}, and finally, the particle is evolved until the end of the timestep with the new momentum. The energy of the particle changes when it collides with the wall. At the end of each timestep, the total change in the particle energies is used to infer the collision induced pressure on the wall and, consequently, its contribution to the acceleration of the wall.

\item The $2\to 2$ collisions between the particles are checked. The input parameter that determines the strength of the self-interactions is a length scale $r_c$, which roughly speaking corresponds to the radius of the hard spheres. In order to avoid checking all particle pairs, the simulation volume is divided into cubic cells of the size $(2r_c)^3$, and the condition for collisions is checked only with the particles in the same and the nearest cells. As particles do not interact at distances greater that $2r_c$, such a cell size is sufficient. If the collision condition is satisfied, the momenta of the two colliding particles are changed so that in the center of mass frame the momenta are back-to-back with a randomly chosen direction. The particles are not moved in the collisions, and collisions of particles at different sides of the wall are not allowed.

\item The bubble radius is evolved with the Euler method using Eq.~\eqref{eq:Reom}. We stop the simulation before the fluid profile that builds up in front of the wall reaches the boundary of the box.
\end{enumerate}

\begin{figure*}
    \includegraphics[width=0.9\textwidth]{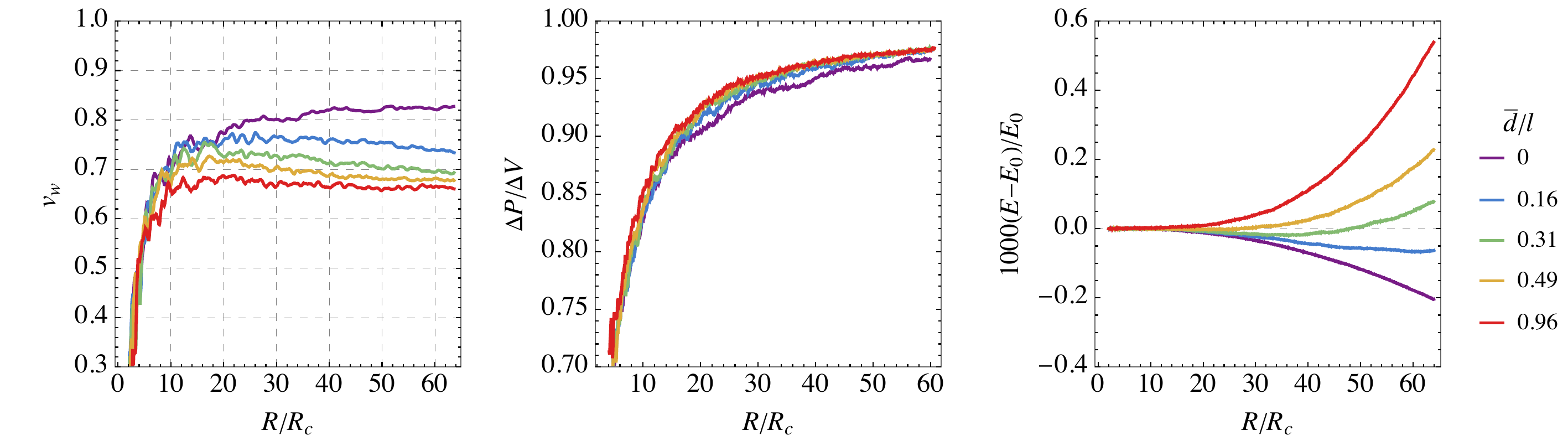}
    \caption{The wall velocity, the pressure difference across the wall and the total energy of the system as a function of the bubble radius for $T_-/m_+ = 0.3$ and different self-interaction strengths indicated by the color coding.}
    \label{fig:Tmpermp03}
\end{figure*}

The length of the time step $\Delta t$ is chosen such that at each time step the distance that particles can travel is much smaller than the box size, $\Delta t = L/1000$, where $L$ is the edge length of the simulation cube. At each timestep, we monitor the total energy of the system and we find that the total energy is conserved in the simulation at permil accuracy, as shown in the right panel of Fig.~\ref{fig:Tmpermp03}. Moreover, we have checked that the probability that the particle interacts with the wall twice within a timestep is negligible, $\mathcal{O}(10^{-8})$, and that the probability that a particle collides more than once within a timestep is relatively small, $\mathcal{O}(10^{-2})$.

\begin{figure}
    \includegraphics[width=0.45\textwidth]{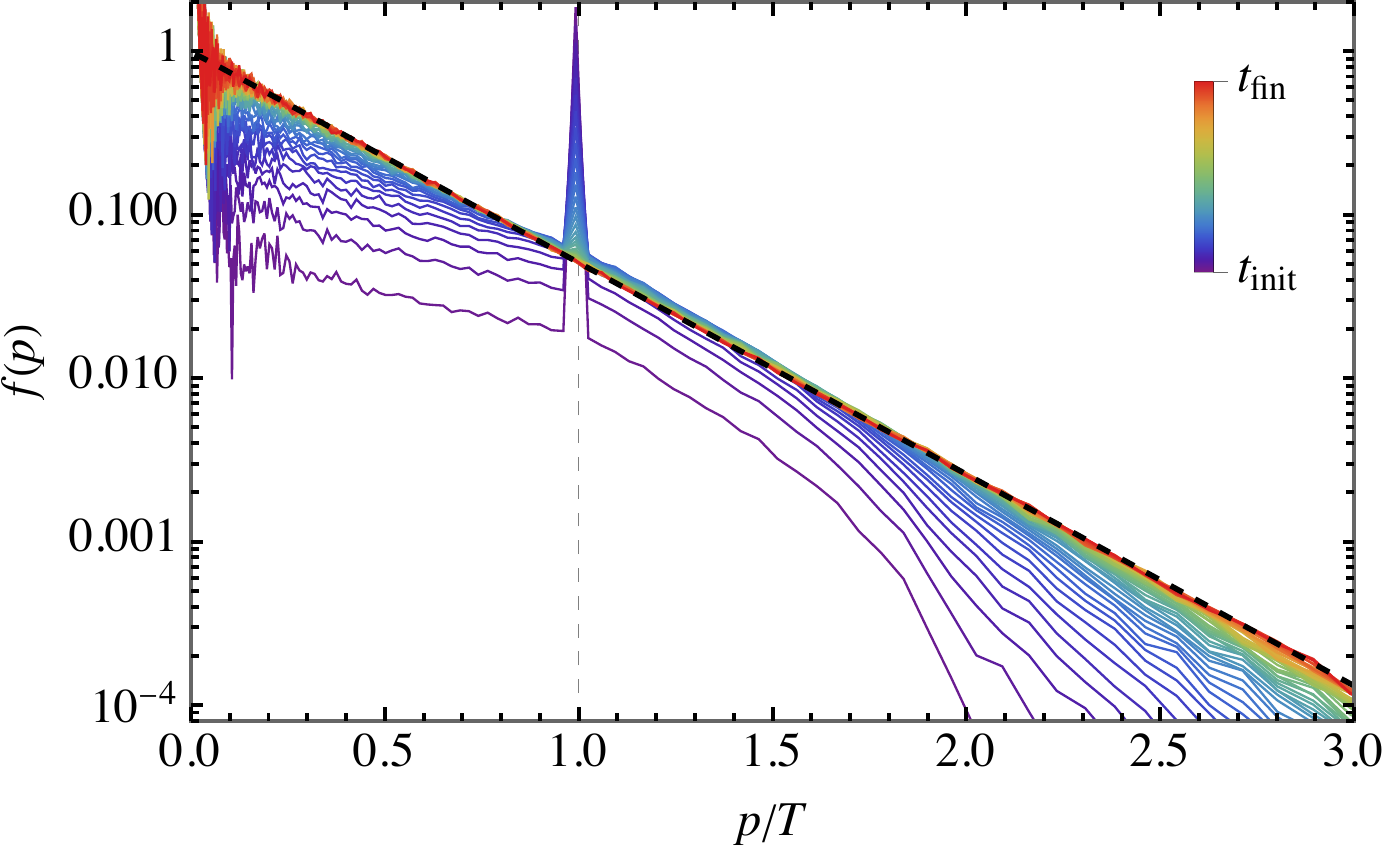}
    \caption{The evolution of the momentum distribution of self-interacting particles starting from a Dirac delta function distribution at $p=T$. The black dashed line shows the Maxwell-Boltzmann distribution.}
    \label{fig:thermalization}
\end{figure}

The particle self-interactions relaxe the momentum distribution of the particles towards the relativistic Maxwell-Boltzmann distribution. In Fig.~\ref{fig:thermalization} we show the momentum distribution of the particles at different times. This simulation does not contain a bubble in the simulation box, and the momentum distribution of the particles is initialized to a Dirac delta function at momentum $p=T$. We see that, due to the momentum exchange between particles in $2\to 2$ collisions, the distribution quickly reaches the Maxwell-Boltzmann distribution.

\begin{figure*}
    \includegraphics[width=0.9\textwidth]{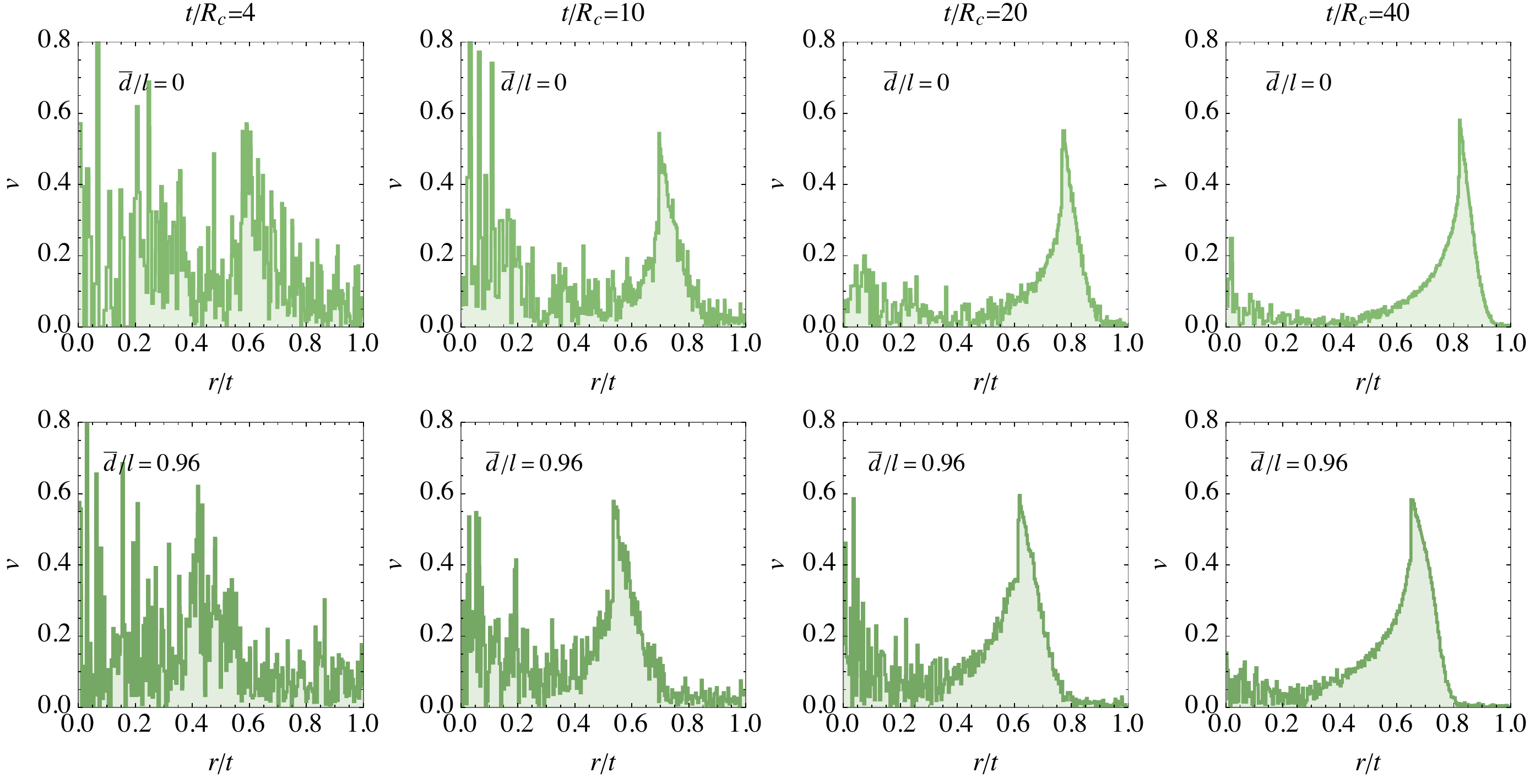}
    \caption{Snapshots of the fluid velocity profile at different times for $T_-/m_+ = 0.3$. The upper and lower panels show the free and self-interacting cases.}
    \label{fig:vprofs}
\end{figure*}

In Fig.~\ref{fig:Tmpermp03} we show the bubble wall velocity $v_w$, the pressure difference across the wall $\Delta P$, and the total energy of the system (particles + bubble) as a function of the bubble radius in a benchmark case with $T_-/m_+ = 0.3$ for self-interaction strengths of the particles. The velocity is averaged over 10 timesteps and the pressure difference over 100 timesteps. The oscillations in these curves reflect particle number in the simulation and we have checked that these oscillations get smaller with increasing particle number. From the right panel, we see that the total energy of the system is conserved at permil accuracy and the violations in the energy conservation increase with increasing self-interaction strength.

In Fig.~\ref{fig:vprofs} we show snapshots of the radial velocity distribution of particles in two of the benchmark cases shown in Fig.~\ref{fig:Tmpermp03}. We see that the distribution at early times includes large uncertainties, which smooth out as the bubble grows.

The simulation runs very quickly if the strength of the self-interactions is small. However, with sizable self-interactions, it can take a relatively long time. For example, the simulation with the strongest self-interactions from which the results are shown in Figs.~\ref{fig:Tmpermp03} and \ref{fig:vprofs}, including $6.4\times 10^6$ particles, took 16 hours on a single core on a Apple M1 Pro, whereas in the case without self-interactions and the same number of particles the running time was only 2.5 minutes. Currently, the code cannot be run in parallel, but, as for any $N$-body simulation, parallelization is possible.

\section{Results}

\subsection{Free particles}

\begin{figure}
    \includegraphics[width=0.45\textwidth]{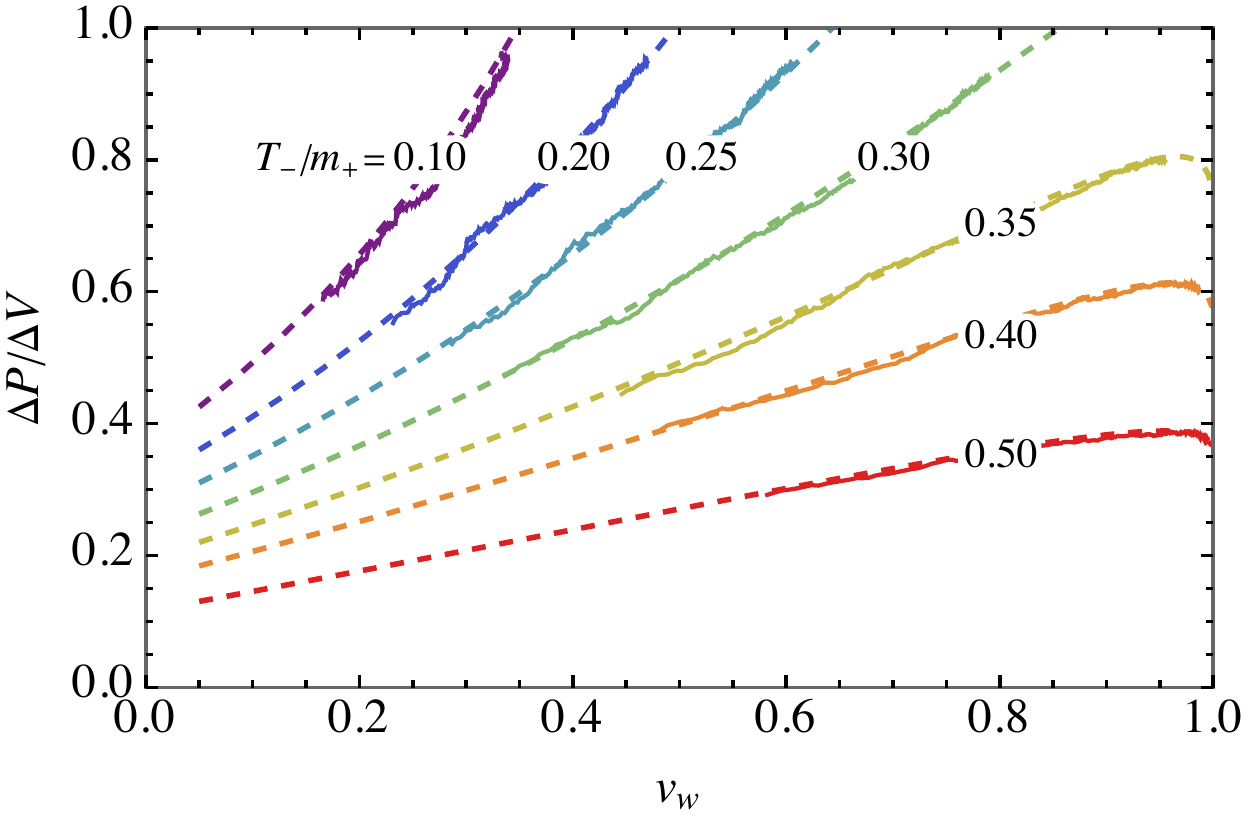}
    \caption{Pressure on the bubble wall caused by free particles for $\alpha = 0.6$ and $T_- \gg m_-$. The dashed contours show the analytical result~\eqref{eq:P} and the solid curves are from numerical simulations.}
    \label{fig:PT}
\end{figure}

The case of free particles is expected to approximate scenarios where the particles' mean free path exceeds the thickness of the fluid shell surrounding the bubble. As shown in Fig.~\ref{fig:PT}, the simulation results match well with analytical estimates obtained from Eq.~\eqref{eq:P} with $f_+(p)=0$. Moreover, we see that $\Delta P$ reaches its maximum at $v_w < 1$. This is because the particles can penetrate the wall if $v_w$ is sufficiently large, and consequently, by Eq.~\eqref{eq:F-}, less energy is transferred away from the wall. For fixed $\alpha$, we further find that $\Delta P/\Delta V \propto m_+^2/T_-^2$ in the relativistic limit. This is expected as $\Delta P \propto m_+^2 T_-^2$ and fixing $\alpha$ implies that $\Delta V \propto T_-^4$.

\begin{figure}[t]
    \includegraphics[width=0.45\textwidth]{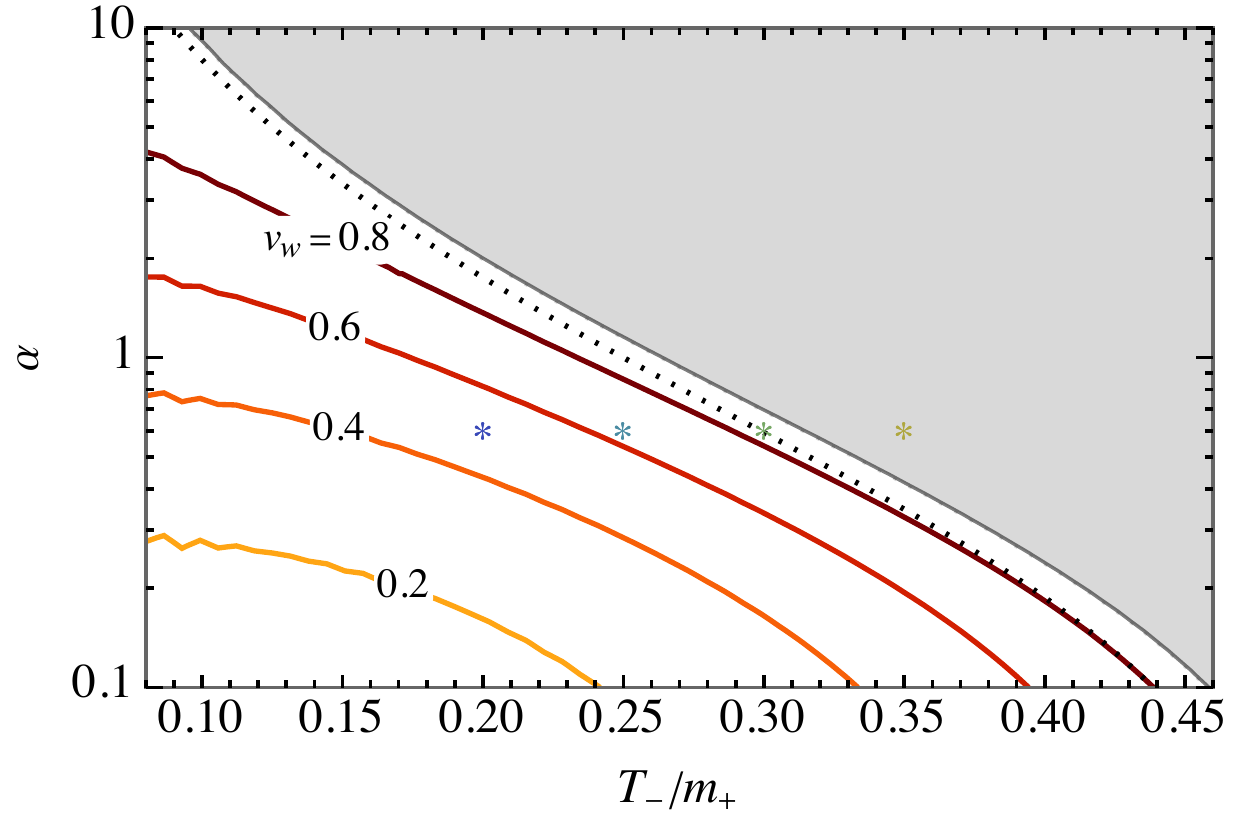}
    \caption{The labelled curves show the terminal bubble wall velocity in a bath of free particles with $f_-(p) = e^{-p/T_-}$ and $f_+(p)=0$ obtained using \eqref{eq:P}. In the gray region, the wall does not reach a terminal velocity. The stars indicate the benchmark cases in Figs.~\ref{fig:vwl} and \ref{fig:rhoprof}. The dashed curve shows the Bodeker\&Moore limit for runaway bubble walls.}
    \label{fig:vterm}
\end{figure}

The terminal velocity can be estimated analytically using Eq.~\eqref{eq:P} with $f_+(p)=0$ by finding the wall velocity $v_w$ for which $\Delta  P = \Delta V$\footnote{This estimate assumes planar walls or large bubbles for which the second term on the r.h.s of Eq.\eqref{eq:Reom} is small.}. As shown in Fig.~\ref{fig:vterm}, the terminal velocity is determined by $\alpha$ and $T_-/m_+$, and increases with both of them. In the gray region, the pressure can never become large enough to stop the wall from accelerating. The last effect can also be understood from Fig.~\ref{fig:PT}, in which $\Delta P/\Delta V < 1$ for the curves with $T_-/m_+ \leq 0.35$ for any $v_w$. For comparison, the dashed curve shows the Bodeker\&Moore result~\cite{Bodeker:2009qy} that, as $\Delta P$ has a maximum at $v_w<1$ while Req.~\cite{Bodeker:2009qy} considered only the $v_w\to 1$ limit of $\Delta P$, slightly underestimates the limiting value of $\alpha$. We note, however, that our analysis includes only $1\to1$ processes at the bubble wall, but, as shown in~\cite{Bodeker:2017cim}, $1\to n$ processes can forbid the runaway behavior even if the pressure arising from the $1\to1$ processes is not large enough.

\subsection{Self-interacting particles}

\begin{figure}[t]
    \includegraphics[width=0.49\textwidth]{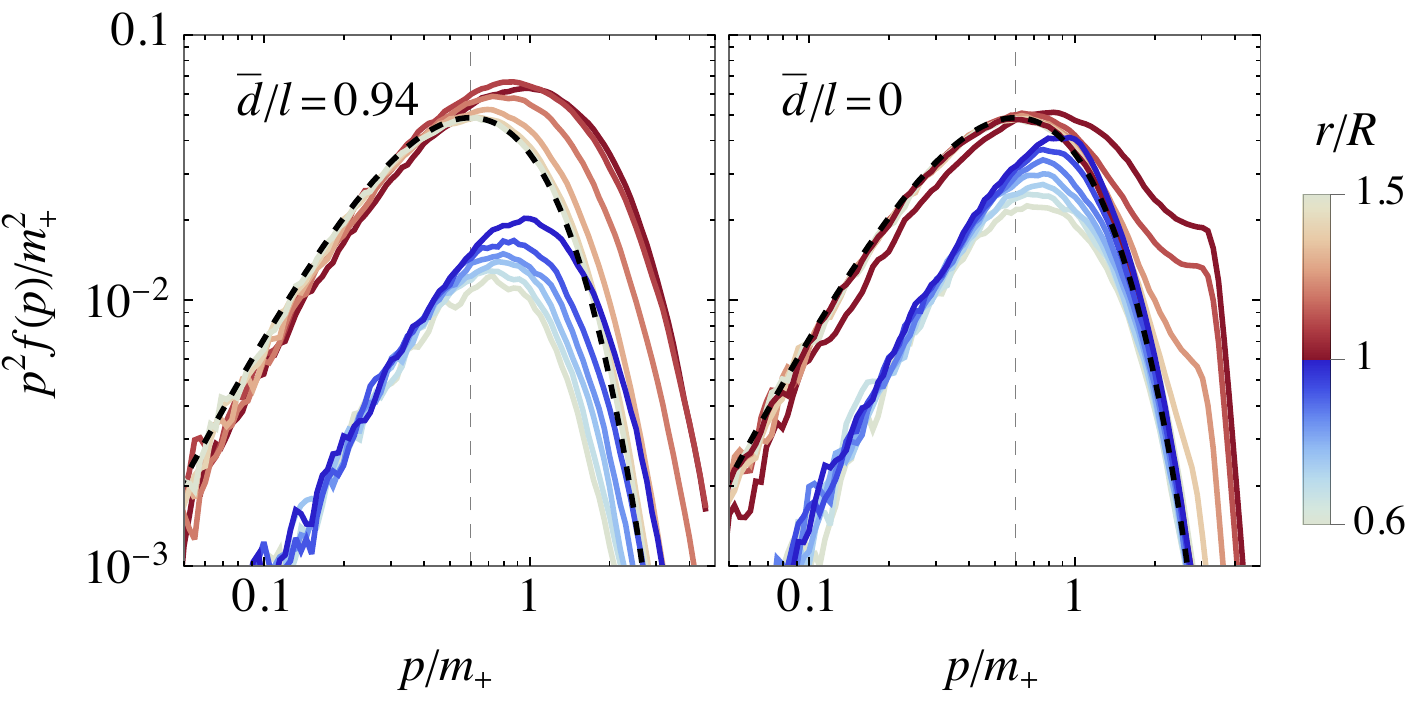}
    \caption{Momentum distribution of particles in spherical shells around the bubble center at radius $r$ indicated by the color coding with (left panel) and without (right panel) self-interactions for $\alpha=0.6$, $T_- = 0.3 m_+$ and $m_- = 0.01m_+$. The the dashed black curve shows $f(p) = e^{-p/T_-}$ and the vertical dashed line highlights $p = 2T_-$.}
    \label{fig:f}
\end{figure}

As shown in the left panel of Fig.~\ref{fig:f}, the self-interactions keep the distribution thermal around the wall. Our simulations conserve the particle number, so the momentum distribution includes a non-vanishing chemical potential around the bubble wall and inside the bubble. As high-momentum particles are more likely to penetrate the wall, the momentum distribution inside the bubble is peaked at higher momentum than outside the bubble. Allowing for strong number changing processes, the chemical potential would remain zero everywhere and the maximum of $p^2 f(p)$ inside the bubble would move towards lower momenta, reflecting the smaller fluid energy density. In the free case, the particles making up the overdensity around the wall move away from it and rise to the high momentum bump in the right panel of Fig.~\ref{fig:f}.

\begin{figure}
    \includegraphics[width=0.4\textwidth]{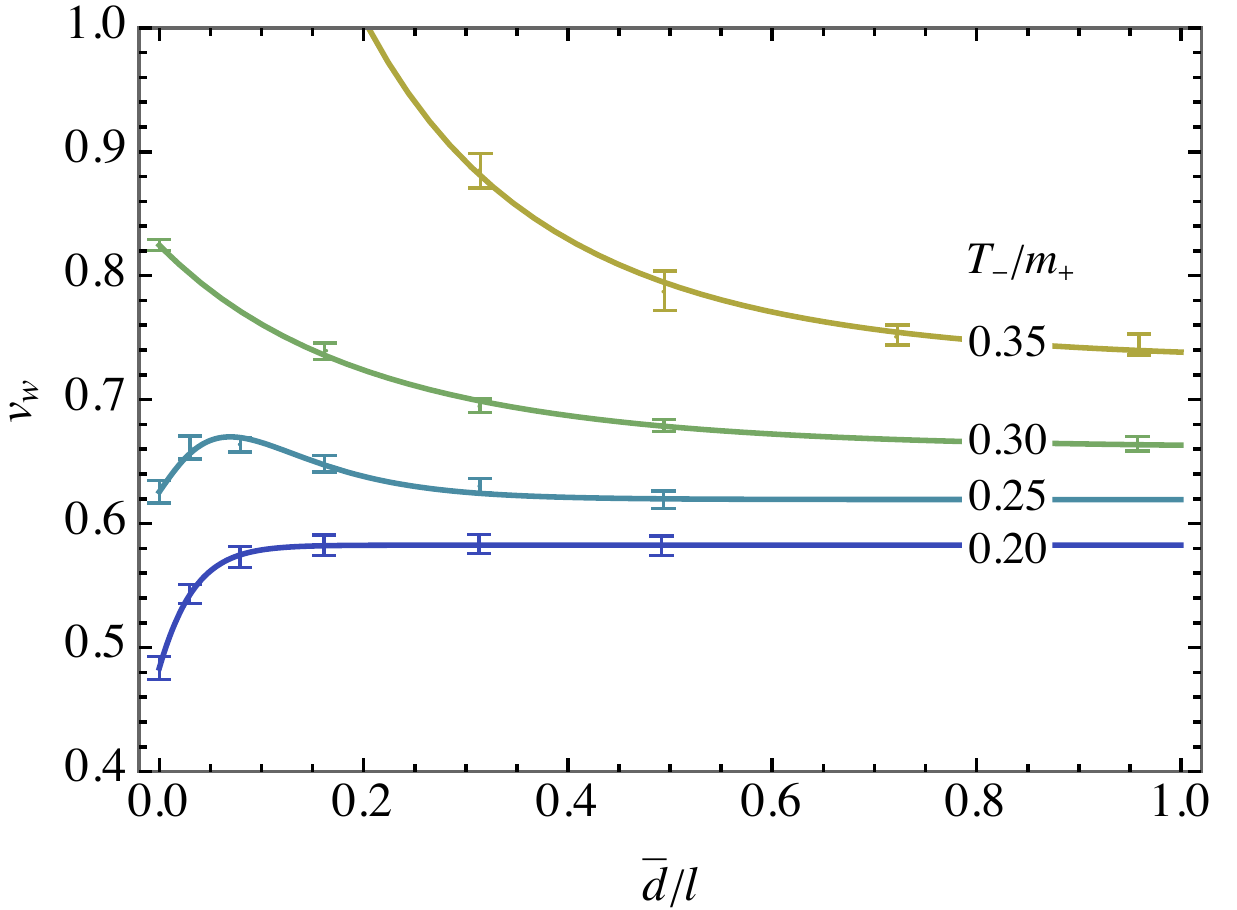}
    \caption{Terminal bubble wall velocity at $R = 64 R_c$ as a function of the inverse of the mean free path $l$ of the particles in the fluid for $\alpha=0.6$ and $m_- = 0.01m_+$. The points with $95\%$ error bars are obtained from the simulations and the solid curves show fitting functions.}
    \label{fig:vwl}
\end{figure}

We compute the particle's mean free path $l$ in terms of their mean separation $\bar{d} \approx n_-^{-1/3}$ by counting the number of collisions per timestep. As shown in Fig.~\ref{fig:vwl}, the terminal velocity reaches a constant value if $l$ is small enough. The numerical results can be approximated well by a 4-parameter fitting function\footnote{The fitting function can be parametrized as $a_1/[(1 + (a_1-1) e^{-a_2 \bar{d}/l})( 1 + a_3 e^{-a_4 \bar{d}/l})]$.} shown with the solid curves. The velocities are shown when $R = 64 R_c$.  We find that for weak self-interactions the velocity of the wall still slowly changes at the end of the simulation, indicating that the wall may always reach the same terminal velocity in infinite time. This is expected as $l$ should be compared to the thickness of the fluid shell around the wall, which increases with the bubble radius.

Fig.~\ref{fig:vwl} shows that the terminal velocity can be either smaller or larger in the self-interacting case than in the free case. Interactions tend to heat the fluid and increase its number density around the wall. The first process results in a decrease in pressure, thus increasing the terminal velocity, while the latter has the opposite effect. Whether the terminal velocity increases or decreases depends on which effect dominates. The simulations indicate that stronger interactions slow down the wall if the wall moves faster than sound, \ie, when $v_w \gtrsim c_{s} \approx 1/\sqrt{3}$, and speed it up otherwise.

\begin{figure}
    \includegraphics[width=0.48\textwidth]{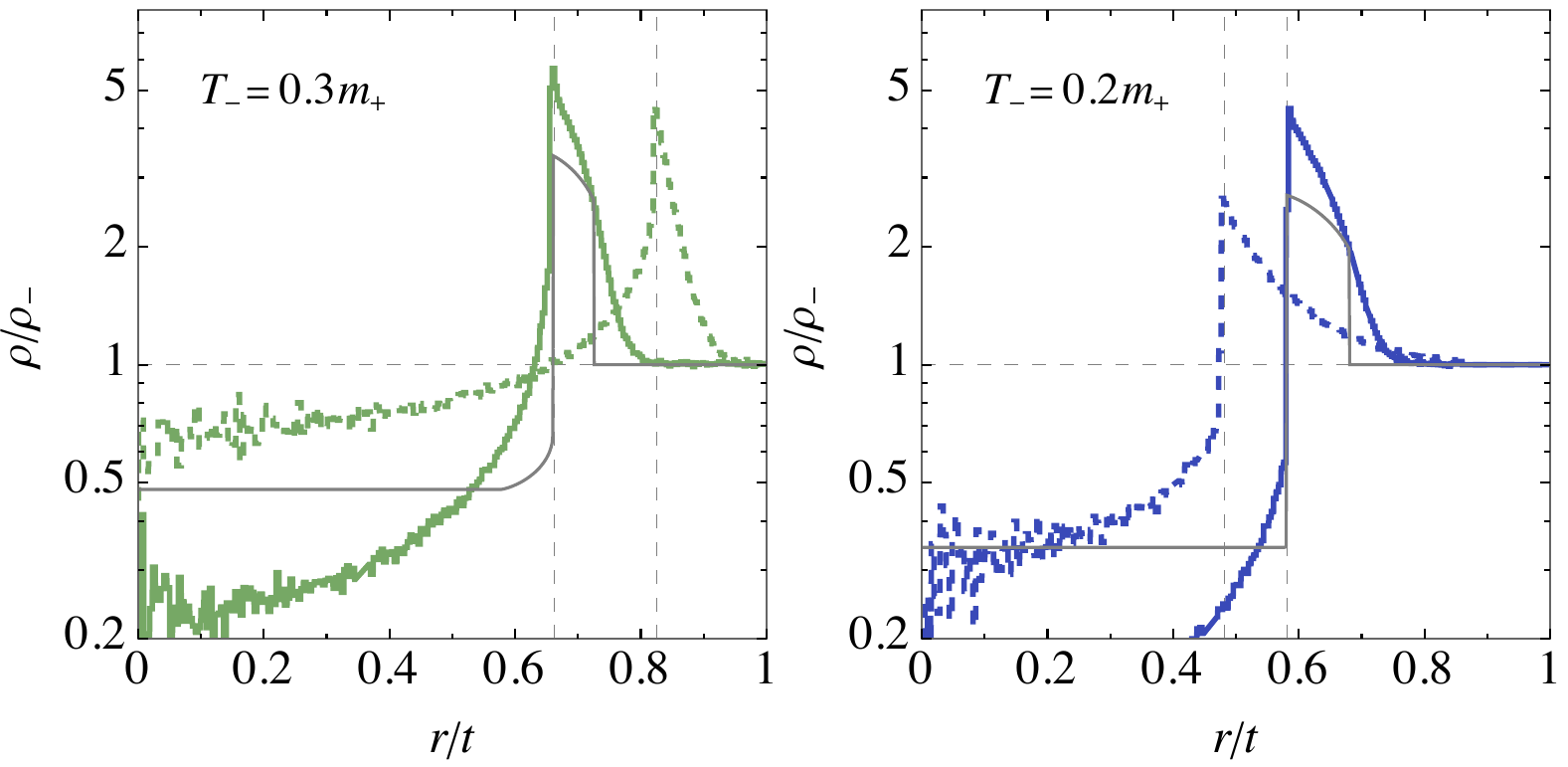}
    \caption{Fluid density profiles with (solid) and without (dashed) self-interactions for $\alpha = 0.6$. The mean free paths in the self-interacting cases are $l=1.0\bar{d}$ (left) and $l=1.8\bar{d}$ (right). The thin gray curves show the profiles obtained by the hydrodynamic approach, and the vertical gray dashed lines the position of the bubble wall.}
    \label{fig:rhoprof}
\end{figure}

Self-interactions make the front of the density profile steeper and decrease the energy density inside the bubble, as is shown in Fig.~\ref{fig:rhoprof}. For comparison, Fig.~\ref{fig:rhoprof} also shows the hydrodynamic profiles for the values of $v_w$ and $\alpha$ corresponding to the self-interacting case. The hydrodynamic shell profile in front of the wall is considerably thinner and lighter. This is likely because we do not simulate particle number-changing processes. Such processes would heat the fluid in front of the wall, and therefore, more particles could penetrate the wall. Since a heavier shell is expected to slow down the wall more than a lighter one, the number-changing processes would also affect the terminal velocity of the wall.

The purpose of simulating particle-particle interactions is to model the effect of the collision term in the Boltzmann equation. Although currently we make several simplifying assumptions about the nature of such interactions, there are no fundamental obstacles for simulating general collision terms. In particular, the approach can be straightforwardly applied to several particle species allowing for example annihilation or bremsstrahlung processes, and interactions whose strength depends on the phase. To model quantum statistics, the collision probability can be adjusted by appropriate Bose enhancement/Pauli blocking factors estimated from the local phase space density in the simulation. Finally, the simulated particles do not have to correspond to individual particles, but to collections of particles analogously to smoothed particle hydrodynamics.

\section{Conclusions}

We have formulated a new particle-based approach to bubble dynamics in fluids. As a proof of concept, we simulated expanding thin-wall bubbles interacting with a single species of particles with conserved particle number. We adopted a framework similar to the so-called ballistic-limit, which in the earlier literature is used to describe very strong transitions, and extended this approach to describe also slower walls. This allowed us to compute the friction exerted on the bubble wall and its final velocity depending on particle masses and their momentum distribution. In particular, in the case in which interactions between particles can be neglected, we have derived these quantities analytically. We have included also particle self-interactions to quantify the thermalization of the surrounding fluid and its effect on bubble dynamics. This allowed us to make contact with the existing results which assume local thermal equilibrium.

The main advantage of this approach is its ability to describe processes without local equilibrium. One example of this comes with strongly supercooled transitions where particles outside the bubble are diluted by expansion and their interactions can be neglected. In such transitions also the profile of the fluid shell behind the wall becomes very sharp and it is not safe to assume that particles there are in equilibrium either. Other examples are cases involving heavy particles out of thermal equilibrium. This is naturally realised in models of filtered dark matter~\cite{Baker:2019ndr,Chway:2019kft} where large mass change upon wall crossing may have a crucial effect on the wall velocity.  

Another key advantage of our method is its ability to treat the wall expansion in dynamical situations. The collapse of false vacuum bubbles at the end of phase transition is one such example, specifically the interactions between the squeezed particles and the wall. This phenomenon has been already described assuming the fluid remains in local equilibrium~\cite{Kurki-Suonio:1995yaf,Cutting:2022zgd}. However, recently many studies try and use squeezing of heavy out of equilibrium particles as a source of black hole formation~\cite{Baker:2021nyl,Baker:2021sno,Huang:2022him} while neglecting the backreaction on the wall. Our method is perfectly suited to model both light and heavy species and check if the force they exert on the wall would not spoil the mechanism.

\begin{acknowledgments}
\vspace{4pt}\noindent\emph{Acknowledgments} -- This work was supported by the Spanish MINECO grants IJC2019-041533-I, FPA2017-88915-P, and SEV-2016-0588, the Spanish MICINN (PID2020-115845GB-I00/AEI/10.13039/501100011033), the grant 2017-SGR-1069 from the Generalitat de Catalunya, the Polish National Science Center grant 2018/31/D/ST2/02048, the Polish National Agency for Academic Exchange within Polish Returns Programme under agreement PPN/PPO/2020/1/00013/U/00001, the Estonian Research Council grants PRG803 and MOBTT5, and the EU through the European Regional Development Fund CoE program TK133 ``The Dark Side of the Universe". IFAE is partially funded by the CERCA program of the Generalitat de Catalunya.
\end{acknowledgments}

\bibliography{gw}

\end{document}